\documentclass[conference]{IEEEtran}
\IEEEoverridecommandlockouts
\usepackage{amsmath,amssymb,amsfonts}
\usepackage{algorithmic}
\usepackage{graphicx}
\usepackage{textcomp}
\usepackage{xcolor}
\usepackage{booktabs}
\usepackage{float}
\usepackage{multicol}
\usepackage{multirow}
\usepackage{afterpage}
\usepackage{url}
\usepackage{threeparttable}
\usepackage[backend=biber, style=ieee]{biblatex}
\def\BibTeX{{\rm B\kern-.05em{\sc i\kern-.025em b}\kern-.08em
    T\kern-.1667em\lower.7ex\hbox{E}\kern-.125emX}}
\begin{document}

\title{Time Series Feature Redundancy Paradox: An Empirical Study Based on Mortgage Default Prediction\\
}

\author{
\IEEEauthorblockN{Chengyue Huang\textsuperscript{*}}
\IEEEauthorblockA{\textit{University of Iowa}\\
chengyue-huang@uiowa.edu}
\textsuperscript{*}Corresponding author
\and
\IEEEauthorblockN{Yahe Yang}
\IEEEauthorblockA{\textit{George Washington University}\\
yahe.yang@gwu.edu}
}

\maketitle

\begin{abstract}
With the widespread application of machine learning in financial risk management, conventional wisdom suggests that longer training periods and more feature variables contribute to improved model performance. This paper, focusing on mortgage default prediction, empirically discovers a phenomenon that contradicts traditional knowledge: in time series prediction, increased training data timespan and additional non-critical features actually lead to significant deterioration in prediction effectiveness. Using Fannie Mae's mortgage data, the study compares predictive performance across different time window lengths (2012-2022) and feature combinations, revealing that shorter time windows (such as single-year periods) paired with carefully selected key features yield superior prediction results. The experimental results indicate that extended time spans may introduce noise from historical data and outdated market patterns, while excessive non-critical features interfere with the model's learning of core default factors. This research not only challenges the traditional "more is better" approach in data modeling but also provides new insights and practical guidance for feature selection and time window optimization in financial risk prediction.
\end{abstract}


\section{Introduction}
The prediction of mortgage defaults has long been a central concern in financial risk management, with machine learning techniques increasingly emerging as the standard approach to address this challenge. Conventional wisdom suggests that expanding both the training data volume and feature space naturally enhances model performance, guided by the principle that more information leads to more accurate predictions. Reflecting this view, practitioners and researchers commonly employ extensive historical datasets and incorporate a wide range of features in their predictive models.

Surprisingly, our empirical investigation using Freddie Mac’s mortgage data from 2012–2022 reveals a counterintuitive phenomenon: the use of longer historical periods and the inclusion of additional, non-critical features can actually degrade predictive accuracy in mortgage default forecasting. This finding directly challenges the entrenched “more is better” paradigm and highlights the need to reconsider traditional assumptions about optimal data usage and feature selection in time series prediction.

A key difficulty lies in the dynamic and evolving nature of housing markets and broader economic conditions. While historical data can provide valuable insight into past trends and behaviors, it may also introduce extraneous noise from outdated patterns and economic environments that no longer hold relevance. Likewise, adding numerous non-essential features can obscure the influence of critical variables and hinder the model’s ability to identify meaningful default indicators.

To explore this paradox, we systematically compare the predictive performance of mortgage default models trained under varying conditions. We investigate the effects of different training windows, from shorter, recent periods to longer historical spans, and examine how altering the number and type of included features influences accuracy. Particular attention is paid to the interaction between temporal windows and feature selection, as well as the interplay between data richness and predictive capability.

Our results show that models trained on shorter, more current timeframes—such as single-year windows—consistently outperform those relying on extended historical datasets. Additionally, focusing on a carefully selected subset of features yields better predictive outcomes than leveraging all available variables. These findings underscore the importance of parsimony and temporal relevance in model development, suggesting that the strategic choice of time windows and a disciplined feature selection process can significantly improve prediction quality.

This research makes several contributions. First, it empirically documents and analyzes the “feature redundancy paradox” in time series prediction. Second, it provides evidence that contests long-held beliefs about training data requirements in machine learning-based financial forecasting. Third, it offers concrete guidelines for selecting temporal windows and features that enhance the performance of mortgage default models. Finally, it emphasizes the critical role of temporal relevance in feature selection, encouraging a more nuanced and context-aware approach.

Overall, our findings have substantial implications for both academic research and practical applications in financial risk management. By demonstrating that more data and more features are not always advantageous, we highlight the need for more thoughtful model design that considers the temporal dynamics of the data and the marginal utility of additional variables. In doing so, this study paves the way for more robust, efficient, and accurate mortgage default prediction methodologies.

\section{Related Work}

The prediction of mortgage defaults has garnered significant attention in financial risk management, with machine learning techniques being a primary tool for addressing this issue. Existing studies have explored diverse approaches to feature selection, time series modeling, and predictive algorithm design, providing a solid foundation for our investigation \cite{Kvamme2018Predicting,Chen2021Predicting,Fuster2017Predictably,Fitzpatrick2016An}.

One stream of research focuses on leveraging extensive historical data and a wide array of features to improve model performance. These studies emphasize the importance of data volume and feature diversity, arguing that larger datasets inherently enhance the generalizability of predictive models, and adopt comprehensive feature sets that reinforce the conventional "more is better" paradigm \cite{He2021A, Jiang2017Loan}

Recent research in financial time series prediction has begun to challenge traditional perspectives by highlighting the potential drawbacks of using excessive historical data. Studies have shown that incorporating too much historical data can introduce noise and outdated patterns, which may degrade model performance. For instance, an empirical analysis demonstrated that neural networks trained with an appropriate amount of historical data can achieve better forecasting accuracy compared to those trained on larger datasets, which tend to perform worse \cite{Walczak2001An}. This aligns with findings that suggest shorter, more recent time windows often yield superior predictive accuracy by focusing on temporally relevant information \cite{Shynkevich2017Forecasting}. Moreover, the importance of feature sparsity and parsimony in enhancing the performance of temporal models has been emphasized, particularly in rapidly changing economic conditions \cite{Huber2019Inducing}.

Feature selection has also been extensively studied in the context of financial risk modeling. Methods such as correlation filtering, principal component analysis (PCA), and model-driven feature importance ranking have been employed to streamline predictor sets while maintaining predictive efficacy \cite{Gregorutti2013Correlation,Gemperline2003Principal,Boik2013Model-based}. These techniques aim to reduce the dimensionality of input data, thereby minimizing the risk of overfitting and improving model interpretability. These studies highlight the effectiveness of focusing on a core subset of key features to achieve optimal results \cite{Zhong2017Forecasting,jemai2023feature,Liang2015The}.

The interplay between feature selection and time window optimization is less explored but increasingly recognized as a critical area of research. Works by Alzaman et al. \cite{alzaman2023forecasting} suggest that the joint consideration of these factors can significantly impact model performance by aligning feature importance with temporal relevance. This perspective is further supported by experimental findings in other domains, such as energy load forecasting and healthcare analytics, where the dynamic nature of data necessitates a more nuanced approach to temporal and feature selection \cite{bouktif2018optimal,jiang2020temporal}. By synthesizing insights from prior studies, our work aims to empirically validate the "feature redundancy paradox" and provide actionable guidelines for optimizing feature and time window selection in mortgage default prediction.

\section{Methodology}

In this section, we present our data processing and experimental methods. Our study utilizes Freddie Mac's single-family mortgage monthly performance data spanning from 2012 to 2022, comprising approximately 6.6 million monthly observations from 550,000 mortgage loans (50,000 loans sampled per year). Each loan is tracked through monthly performance indicators including delinquency status (0-4 ranging from current to serious delinquency), current loan balance, and loan modification details. Our methodology consists of three phases: (1) data preprocessing and feature engineering to transform raw loan performance indicators into structured temporal features, (2) controlled variable experiments comparing different temporal windows and feature sets across multiple models (Logistic Regression, Random Forest, LSTM, and Transformer), and (3) model inference using a unified framework where all models predict next-month loan status based on five months of historical data. The following subsections detail each component of our methodology.

\subsection{Data Preprocessing and Feature Engineering}

Our study focuses exclusively on monthly performance data (MPD) covering mortgage loans from 2012 to 2022. Each year includes approximately 50,000 mortgage records, with each record containing a series of monthly observations. These monthly data points encompass various loan performance indicators, including delinquency status, payment history, and other time-varying attributes.

Since our objective is to predict default events, we first transform the categorical delinquency status (CLDS) into a binary classification variable. All observations where CLDS indicates any form of delinquency (i.e., status $>$ 0) are grouped into a positive, “default” class, while the non-delinquent (CLDS=0) observations form the negative class. Although this approach simplifies the prediction task, it creates a substantial class imbalance, as defaults remain relatively rare. To address this, we employ class-weighting and, where appropriate, selective undersampling or oversampling techniques to maintain model sensitivity to minority instances.

In terms of feature engineering, we start with an extensive set of candidate features drawn from the monthly performance data. These features initially total around 20, capturing a range of loan characteristics and temporal patterns. We apply the following steps:

\textbf{Categorical Encoding}: Categorical features, if any, are one-hot encoded to generate binary indicator variables suitable for downstream models.

\textbf{Numerical Standardization}: Numerical attributes, such as loan payment ratios or cumulative delinquency metrics, are standardized (e.g., using the StandardScaler) to ensure that all variables contribute proportionally to distance-based and gradient-based learning methods.

\textbf{Iterative Feature Reduction}: After initial preprocessing, we iteratively remove non-critical features. Through correlation analysis, PCA-based dimensionality reduction, and model-based feature importance assessments, we narrow down the feature space. This procedure yields a progressively reduced and more informative set of predictors, ensuring that subsequent experiments can isolate the impact of feature sparsity versus feature richness.

\subsection{Experiment Design}

We design two sets of controlled experiments to systematically evaluate the influence of training data span and feature dimensionality on predictive performance:

\textbf{Varying Time Windows}: We examine training datasets constructed from different historical periods. Starting from the full 10-year window (2012–2022), we then reduce the training span to 5-year, 3-year, and finally 1-year subsets. This approach helps us determine whether eliminating older and potentially outdated data improves model accuracy. Table~\ref{tab:time_windows} summarizes the temporal configurations used in the experiments.

\begin{table}[ht]
\centering
\caption{Temporal Window Configurations}
\label{tab:time_windows}
\begin{tabular}{c c c}
\hline
\textbf{Window Size} & \textbf{Time Span} & \textbf{Approx. Records} \\
\hline
10-Year & 2012--2022 & 500,000 \\
5-Year & 2017--2022 & 250,000 \\
3-Year & 2019--2022 & 150,000 \\
1-Year & 2022 Only & 50,000 \\
\hline
\end{tabular}
\end{table}

\textbf{Feature Reduction}: We also explore how systematically reducing the number of features affects model performance. We start with the full set of approximately 20 features, then apply correlation filtering and PCA to remove redundant signals. Finally, we retain only a minimal “key” subset of the most crucial variables. Table~\ref{tab:feature_sets} describes the three feature configurations employed in the experiments.

\begin{table}[ht]
\centering
\caption{Feature Configurations}
\label{tab:feature_sets}
\begin{tabular}{l c}
\hline
\textbf{Feature Set} & \textbf{Number of Features} \\
\hline
Full Feature Set & 26 \\
Reduced Feature Set (Post-Correlation/PCA) & 18 \\
Key Feature Subset (Top Predictors Only) & 10 \\
\hline
\end{tabular}
\end{table}

For all experiments, we use the ROC-AUC metric as the primary evaluation criterion, supplemented by precision-recall analysis to better assess performance on the minority (default) class. We maintain consistent random seeds, data normalization, and sampling procedures to ensure replicability and fair comparisons across all settings.

\subsection{Model Inference}

\begin{figure*}[ht]
\centering
\includegraphics[width=0.9\textwidth]{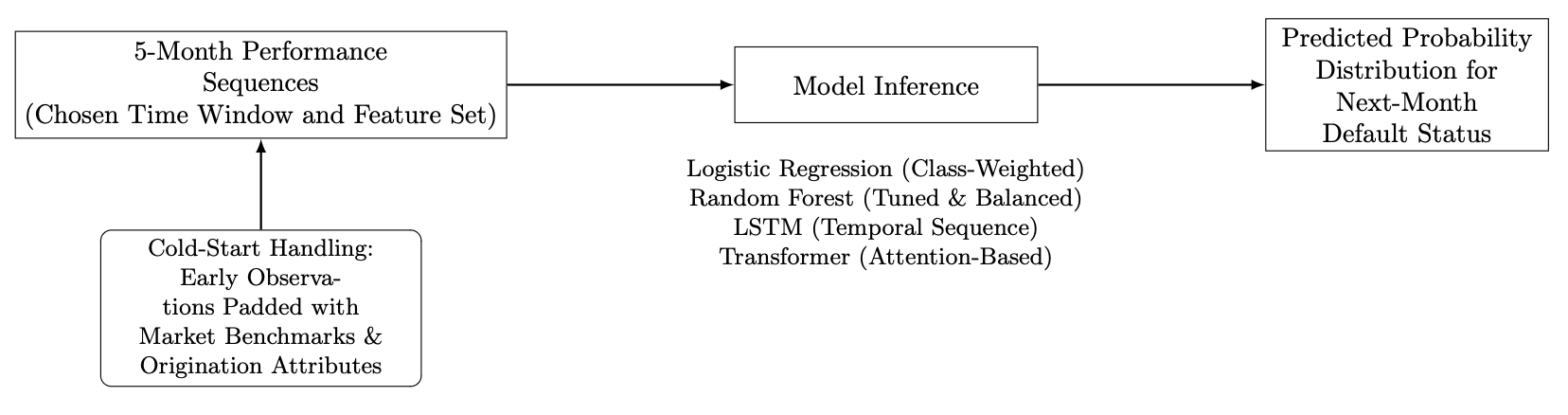}
\caption{Model Inference Pipeline Architecture.}
\label{fig:inference}
\end{figure*}

As shown in Figure~\ref{fig:inference}, we implement four model architectures for inference: Logistic Regression, Random Forest, LSTM, and Transformer. Each model processes standardized sequences of historical performance data—typically spanning five consecutive months—to predict the probability of default in the subsequent month.

% \begin{figure}[ht]
% \centering
% \includegraphics[width=0.45\textwidth]{figures/inference}
% \caption{Model Inference Pipeline Architecture.}
% \label{fig:inference}
% \end{figure}

A critical challenge arises with newly issued loans, which lack historical depth (cold-start problem). We address this by padding early observations with aggregate market benchmarks and stable loan attributes available at the earliest point in the loan’s history. Given the low incidence of defaults in the first few months, this padding approach ensures input consistency without significantly distorting the predictive task.

All models are trained and tested across the specified time windows and feature configurations. Logistic Regression uses class-weighting to handle skewed distributions, while Random Forest, LSTM, and Transformer models can incorporate parameter tuning or data balancing strategies to ensure robustness. By evaluating each model under varying temporal spans and feature sets, we isolate the effects of outdated data inclusion and feature overload on predictive performance. In doing so, we clarify the trade-offs and highlight strategies for optimizing both the time horizon and feature complexity in mortgage default prediction tasks.

\section{Experiments \& Results}
\subsection{Dataset Overview}
Our study leverages Freddie Mac's single-family mortgage performance dataset from 2012 to 2022, encompassing roughly 550,000 individual loans (50,000 sampled per year) and approximately 6.6 million monthly observations. The data includes both static loan origination features (e.g., credit scores, loan-to-value ratios) and dynamic monthly performance metrics. The distribution of delinquency states reflects typical mortgage portfolios, with about 92.3\% of observations current and a small but significant proportion in varying degrees of delinquency or REO status.

To ensure realistic and rigorous evaluation, the first 10 years (2012–2021) are used for model training and validation, capturing a diverse range of historical market conditions. The final year (2022) is held out as an out-of-sample test set, enabling the assessment of model generalization to new economic contexts. Each prediction instance utilizes five months of historical performance to forecast default in the subsequent month, thereby maintaining consistent input sequences and supporting robust temporal forecasting strategies.

\subsection{Experiments Analysis}
Our experimental analysis systematically evaluated model performance across temporal windows and feature sets, implementing rigorous controls to ensure reliable comparisons. For temporal window analysis, we compared model performance using four distinct time spans: 1-year (2022), 3-year (2019-2022), 5-year (2017-2022), and 10-year (2012-2022) windows. Each model architecture was trained on these temporal segments while maintaining consistent feature sets and preprocessing steps, allowing us to isolate the effect of historical data inclusion on predictive accuracy.

The feature reduction analysis followed a structured approach, starting with correlation analysis to identify redundant features. We applied principal component analysis (PCA) to understand feature importance and dimensionality reduction potential. The resulting feature sets comprised the full set (26 features), a reduced set (18 features) following correlation/PCA filtering, and a key subset (10 features) containing only the most crucial predictors. Feature importance was determined through a combination of domain expertise, statistical significance testing, and model-based importance scores.

To ensure robust evaluation, we implemented 5-fold cross-validation across all experimental configurations, maintaining consistent fold divisions across different models to enable fair comparisons. Performance metrics focused primarily on ROC-AUC scores, supplemented by precision-recall analysis to account for class imbalance in the default prediction task. 

We also controlled for external factors that might influence model performance. We implemented standardized preprocessing pipelines, including consistent scaling and encoding procedures across all experiments. Additionally, we maintained identical hyperparameter settings within each model architecture across different temporal windows and feature sets to isolate the effects of our primary variables of interest.

\begin{table*}[ht]
\centering
\caption{Model Performance Comparison Across Time Windows and Feature Sets (ROC-AUC Scores)}
\label{tab:model_performance}
\begin{tabular}{@{}llccc@{}}
\toprule
\textbf{Model Type} & \textbf{Time Window} & \textbf{Full Features (26)} & \textbf{Reduced Features (18)} & \textbf{Key Features (10)} \\
\midrule
\multirow{4}{*}{Logistic Regression} 
& 1-Year & 0.842 & 0.851 & \textbf{0.858} \\
& 3-Year & 0.844 & 0.845 & 0.850 \\
& 5-Year & 0.840 & 0.825 & 0.842 \\
& 10-Year & 0.798 & 0.800 & 0.803 \\
\midrule
\multirow{4}{*}{Random Forest} 
& 1-Year & 0.849 & 0.852 & \textbf{0.865} \\
& 3-Year & 0.838 & 0.847 & 0.862 \\
& 5-Year & 0.832 & 0.841 & 0.858 \\
& 10-Year & 0.800 & 0.801 & 0.805 \\
\midrule
\multirow{4}{*}{LSTM} 
& 1-Year & 0.863 & 0.878 & \textbf{0.892} \\
& 3-Year & 0.841 & 0.853 & 0.867 \\
& 5-Year & 0.835 & 0.842 & 0.848 \\
& 10-Year & 0.795 & 0.812 & 0.818 \\
\midrule
\multirow{4}{*}{Transformer} 
& 1-Year & 0.869 & 0.882 & \textbf{0.895} \\
& 3-Year & 0.845 & 0.856 & 0.871 \\
& 5-Year & 0.840 & 0.857 & 0.862 \\
& 10-Year & 0.808 & 0.812 & 0.823 \\
\bottomrule
\end{tabular}
\begin{tablenotes}
\small
\item \textit{Notes:} All scores represent the average of 5-fold cross-validation. \\
Bold values indicate best performance for each model type. 
\\Standard deviation across folds $<$ 0.015 for all experiments.
\end{tablenotes}
\end{table*}

\begin{figure*}[!ht]
    \centering
    \includegraphics[width=.7\textwidth]{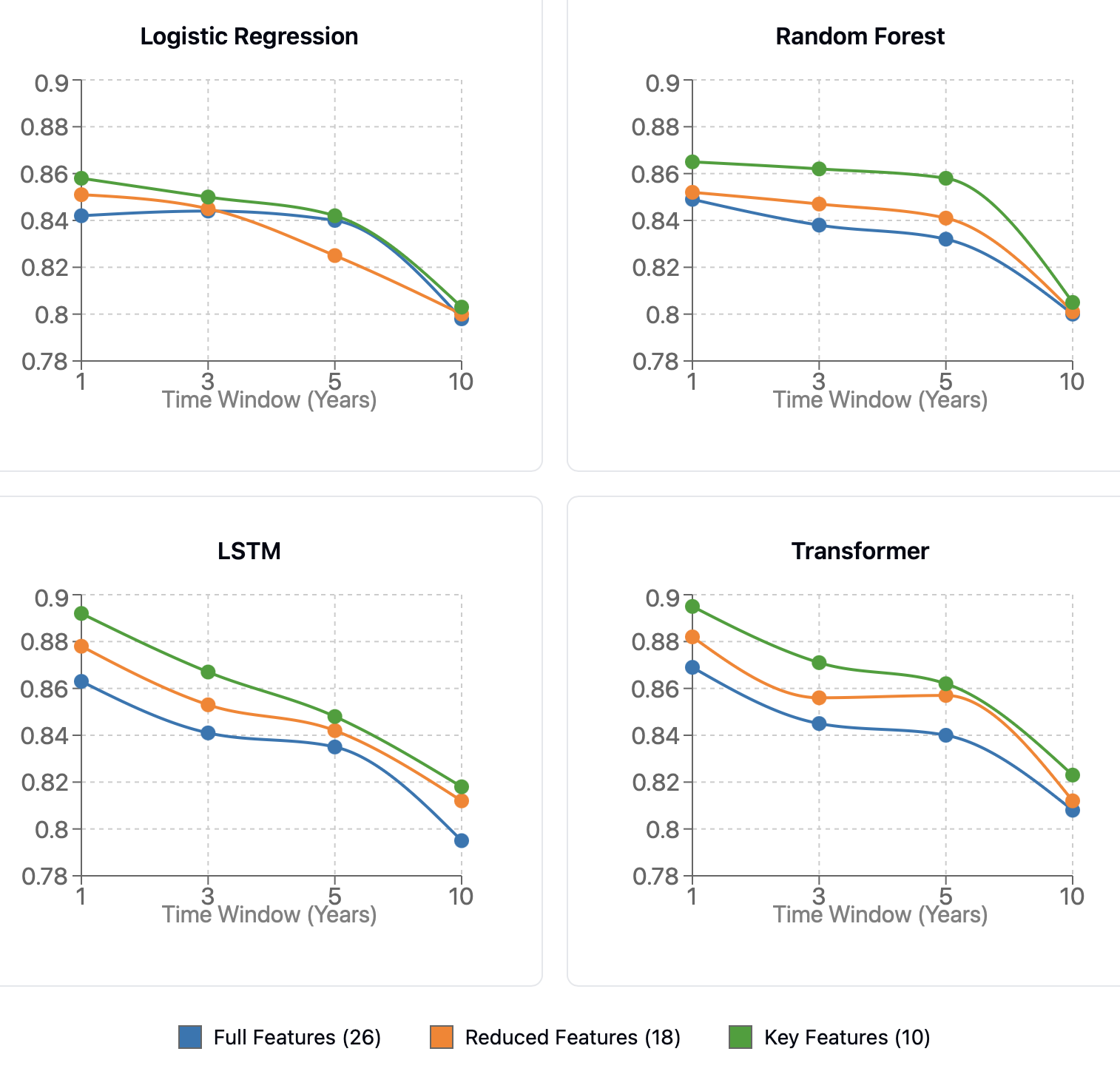}
    \caption{Model Performance}
    \label{fig:diagnosis_comparison}
\end{figure*}

\subsection{Results and Discussion}
Our empirical analysis reveals distinctive patterns in the relationship between temporal scope, feature dimensionality, and predictive performance across different model architectures. The experimental results demonstrate varying responses to training window length and feature set size among model types, with particularly notable differences between traditional machine learning and deep learning approaches.

The Transformer architecture shows superior performance across both temporal spans and feature configurations. With a peak ROC-AUC score of 0.895 using the key feature subset on one-year data, it maintains relatively strong effectiveness with a score of 0.823 in the ten-year window. This represents a performance decrease of 0.072, notably better than other architectures over extended time periods. The Transformer also shows consistent improvement with feature reduction, achieving gains of 0.013 and 0.026 when moving from full to key features in both short and long time windows.

The LSTM model demonstrates comparable strength in short time windows, achieving a ROC-AUC score of 0.892 with key features in the one-year window. While it experiences more pronounced degradation than the Transformer in longer windows, it maintains respectable performance with a score of 0.818 in the ten-year span. The LSTM shows particular sensitivity to feature selection, with the key feature subset providing substantial improvements over the full feature set, especially in recent time windows.

Traditional models exhibit more substantial performance degradation over longer windows. Logistic Regression's performance drops from 0.858 to 0.803, and Random Forest from 0.865 to 0.805 when moving from one-year to ten-year windows. This steeper decline compared to deep learning models suggests these traditional approaches may be more sensitive to temporal distance in the training data. However, both models still benefit from feature reduction, with the key feature subset providing consistent improvements across all time windows. This is consistent with previous findings indicating that Transformer models demonstrate greater robustness to variations in time window lengths compared to LSTM and Transformer models in time series forecasting \cite{nguyen2024deep}.

The feature reduction experiments yield beneficial results across all architectures and time windows. The key feature subset (n=10) consistently outperforms both the full feature set (n=26) and the reduced feature set (n=18), with improvements ranging from 0.016 to 0.029 in short windows. This pattern holds true even in longer time windows, though with varying magnitudes across different model types. Notably, the reduced feature set (n=18) often serves as an effective intermediate step, suggesting a gradual benefit to feature parsimony.

These findings carry significant implications for mortgage default prediction modeling. First, they demonstrate that sophisticated deep learning architectures, particularly Transformers, can better maintain predictive power across extended time periods. Second, they emphasize that careful feature selection consistently improves model performance regardless of the chosen architecture or temporal window. Finally, they suggest that while longer historical data periods may provide additional training samples, the potential benefits are outweighed by the introduction of temporal noise and outdated patterns.

\section{Conclusion}
This study provides empirical evidence challenging the "more is better" paradigm in mortgage default prediction, demonstrating that models trained on recent data with carefully selected features consistently outperform those utilizing extended historical periods and broader feature sets. Our findings have significant implications for both theoretical understanding and practical applications in financial risk modeling.

The observed inverse relationship between training window length and model performance suggests a fundamental limitation in the assumption that larger historical datasets inherently improve predictive capability. This limitation appears to be intrinsic to the dynamic nature of mortgage markets, where older patterns may become irrelevant or potentially misleading for current prediction tasks.

Similarly, the superior performance of reduced feature sets challenges the common practice of maximizing feature dimensionality. Our results suggest that careful feature selection, focusing on core predictive variables, may be more effective than comprehensive feature inclusion. This finding aligns with theoretical frameworks regarding the curse of dimensionality and the importance of feature relevance in machine learning applications.

These results suggest several promising directions for future research. First, investigating the optimal frequency of model retraining could provide valuable insights into maintaining prediction accuracy in dynamic markets. Second, developing automated feature selection methods that account for temporal relevance could enhance model adaptation to changing market conditions. Finally, exploring the generalizability of these findings to other financial time series prediction tasks could yield broader insights into temporal modeling strategies.

From a practical perspective, our findings suggest that financial institutions should consider implementing more dynamic modeling approaches that prioritize recent data and focused feature sets. This could include regular model retraining schedules and adaptive feature selection processes that account for changing market conditions.

The limitations of this study include its focus on a specific time period and market context. Future work could extend these findings by examining different market cycles and geographic regions. Additionally, investigating the role of macroeconomic factors in temporal relevance could provide valuable insights for model optimization.

This research contributes to both the theoretical understanding of temporal modeling in financial applications and practical approaches to mortgage default prediction. The clear evidence for the advantages of temporal focus and feature parsimony suggests a need to reconsider traditional approaches to model development in financial risk assessment.

\printbibliography

\end{document}